# A comparison between the two lobes of comet 67P / Churyumov-Gerasimenko based on D/H ratios in $H_2O$ measured with the Rosetta / ROSINA DFMS

**Isaac R.H.G. Schroeder I** (1), Kathrin Altwegg (1), Hans Balsiger (1), Jean-Jacques Berthelier (3), Michael R. Combi (8), Johan De Keyser (4), Björn Fiethe (5), Stephen A. Fuselier (6,7), Tamas I. Gombosi (8), Kenneth C. Hansen (8), Martin Rubin (1), Yinsi Shou (8), Valeriy M. Tenishev (8), Thierry Sémon (1), Susanne F. Wampfler (2), Peter Wurz (1,2)

(1) Physikalisches Institut, University of Bern, Sidlerstrasse 5, CH-3012 Bern, Switzerland
(2) Center for Space and Habitability, University of Bern, Gesellschaftsstrasse 6, CH-3012 Bern, Switzerland
(3) LATMOS, 4 Avenue de Neptune, F-94100 Saint-Maur, France
(4) Royal Belgian Institute for Space Aeronomy (BIRA-IASB), Ringlaan 3, B-1180, Brussels, Belgium
(5) Institute of Computer and Network Engineering (IDA), TU Braunschweig, Hans-Sommer-Straße 66, D-38106 Braunschweig, Germany
(6) Space Science Division, Southwest Research Institute, 6220 Culebra Road, San Antonio, TX 78228, USA
(7) University of Texas at San Antonio, San Antonio, TX
(8) Department of Climate and Space Sciences and Engineering, University of Michigan, 2455 Hayward, Ann Arbor, MI 48109, USA

## Abstract

The nucleus of the Jupiter-family comet 67P / Churyumov-Gerasimenko was discovered to be bi-lobate in shape when the European Space Agency spacecraft Rosetta first approached it in July 2014. The bi-lobate structure of the cometary nucleus has led to much discussion regarding the possible manner of its formation and on how the composition of each lobe might compare with that of the other. During its two-year-long mission from 2014 to 2016, Rosetta remained in close proximity to 67P / Churyumov-Gerasimenko, studying its coma and nucleus in situ. Based on lobe-specific measurements of HDO and $H_2O$ performed with the ROSINA DFMS mass spectrometer on board Rosetta, the Deuterium-to-Hydrogen ratios in water from the two lobes could be compared. No appreciable difference was observed, suggesting that both lobes formed in the same region and are homogeneous in their Deuterium-to-Hydrogen ratios.

## 1. Introduction

The European Space Agency (ESA) spacecraft Rosetta rendezvoused with the Jupiter-family comet (JFC) designated 67P / Churyumov-Gerasimenko (hereafter 67P) in August 2014, at a distance of about 3.5 AU from the Sun and as the comet was approaching its perihelion. Rosetta remained thereafter in close proximity to the cometary nucleus until September 2016, accompanying 67P to its perihelion at 1.24 AU and then back out again to a heliocentric distance of 3.5 AU. Over the course of its two-year mission, Rosetta studied the coma and nucleus of 67P in situ with a wide array of scientific instruments.

One such instrument, the OSIRIS camera experiment (Keller et al. 2007), made the initial observation of the bi-lobate structure of 67P's nucleus when Rosetta first approached the comet in July 2014. The finding that 67P consists of two lobes, one slightly larger than the other, was a major surprise. Such an interesting shape immediately raised the question of how it came to be – whether it was formed from a single parent or the merger of two (Nesvorný et al. 2018; Massironi et al. 2015), whether the lobes were homogeneous or heterogeneous and, if the lobes had formed separately, whether they had formed in the same region or in different regions before merger (Rickman et al. 2015).

Another of the instrument packages on board Rosetta was the Rosetta Orbiter Spectrometer for Ion and Neutral Analysis (ROSINA), which included a Double Focusing Mass Spectrometer (DFMS) capable of measuring isotopic abundances in the volatile species in the cometary coma (Balsiger et al. 2007). The ROSINA DFMS was almost always active after Rosetta's rendezvous with 67P, continuously analysing the comet's gaseous atmosphere close to its nucleus and usually switched off only during spacecraft manoeuvres to avoid contamination from thruster exhaust. It could identify trace amounts of rare species and isotopologues alongside much more common ones due to its high mass resolution, dynamic range and sensitivity (Hässig et al. 2013), and was used to measure the Deuterium-to-Hydrogen (D/H) ratio in 67P's cometary water by Altwegg et al. (2014, 2017), who found a value of $(5.3 \pm 0.7) \times 10^{-4}$. This is more than three times the terrestrial VSMOW value of $1.6 \times 10^{-4}$ and was an important result for the discussion on the origins of terrestrial oceans. It was also one of the highest D/H ratios yet measured in a JFC.

However, the previously measured cometary D/H ratio (Altwegg et al. 2014, 2017) was a global result for 67P. As there was insufficient spatial information available then for lobe-specific measurements (Altwegg et al. 2017), the possibility of there being some inhomogeneity between the two lobes could not be completely excluded at the time.

D/H ratios can vary drastically amongst the different bodies in the solar system (Altwegg et al. 2014). It is estimated to have been $(2.1 \pm 0.5) \times 10^{-5}$ for the protosolar nebular (PSN) (Levison et al. 2010; Mahaffy et al. 1998), which is comparable to the interstellar value of $(2.0 \sim 2.3) \times 10^{-5}$ derived from solar wind measurements (Geiss & Gloeckler 1998), though this varies by up to a factor of four in the global interstellar medium (Linsky et al. 2006). Relative to the PSN, however, most solar system objects are enriched in deuterium (Altwegg et al. 2014). A possible explanation for this is deuterium-rich ice from the presolar cloud falling into the protosolar nebula (Ceccarelli et al. 2014) and then partially vaporising and recondensing. Subsequent isotopic exchange with molecular hydrogen (Visser et al. 2011) would have favoured the accumulation of deuterium in HDO at low temperatures. As this reaction cannot proceed after water has crystallised, the D/H ratio of a particular body reflects the local values at the time and region where its material condensed (Geiss & Reeves 1981). An alternative explanation – grain-surface chemistry in the cold outer regions of the PSN – has also been proposed as a mechanism for deuterium fractionation (Kavelaars et al. 2011; Furuya et al. 2013). Whichever the case, deuterium enrichment increases radially with heliocentric distance (Geiss & Reeves 1981; Kavelaars et al. 2011; Furuya et al. 2013). Cometary D/H ratios therefore depend on their region of formation.

The two main families of comets were once believed to have formed in very different regions, between Uranus and Neptune for Oort cloud comets (OCCs) (Albertsson et al. 2014) and in the Kuiper belt much further away for Jupiter-family comets (JFCs) (Duncan & Levison 1997). However, several recent measurements have found some JFCs to possess D/H ratios lower than most OCCs (Hartogh et al. 2011; Ceccarelli et al. 2014) and an OCC with a very high D/H ratio (Biver et al. 2016). This showed that D/H ratios in both JFCs and OCCs can vary considerably from comet to comet, which indicates that they formed over a wide range of heliocentric distances and further suggests that their formation regions may overlap (Brasser & Morbidelli 2013). New research by Lis et al. (2019) has even found a correlation between a comet's size, level of activity and D/H ratio. Given that the two lobes of 67P are of comparable but not the same size and that they are most likely a contact binary formed of two separate objects, it is only natural to consider whether or not their D/H ratios might differ, which would imply formation in different regions.

In this work, we compare the D/H ratio in water from 67P's larger lobe with that from its smaller lobe by utilising ROSINA DFMS measurements of HDO and $H_2O$ from the cometary coma in tandem with a recently developed outgassing model (Combi et al. 2019) that can estimate the amount of water contributed by each individual lobe to the gases sampled by Rosetta at any given time.

## 2. Instrumentation & Methodology

The ROSINA DFMS is a double focusing mass spectrometer with a high mass resolution of $m/\Delta m \sim 3000$ at 1% peak height (Balsiger et al. 2007). Its main detector, the MCP **/** LEDA, is a position-sensitive imaging detector comprised of two Micro-Channel Plates (MCPs) in a Chevron configuration, followed by two independent rows (Rows A and B) of 512 anodes on a Linear Electron Detector Array (LEDA). A more detailed description of the instrument may be found in Schroeder et al. (2019) and the references therein.

The overall gain (degree of amplification) produced by the MCP depends on which of the 16 predefined voltage settings (gain steps) is applied. However, the gain corresponding to each gain step changed over time as the detector aged. This had to be corrected for in all DFMS measurements. An additional flat-field correction known as the "pixel gain" was also necessary, due to the non-uniform degradation of the 512 LEDA anodes (pixels) caused by uneven usage of the MCP. The appendix of Schroeder et al. (2019) contains a full description of these corrections and how they are applied.

Fig. 1 shows a shape model of 67P. As a crude first approximation, we can take the small lobe to be the part that lies between -60° and +60° longitude, and the large lobe to be the rest. Assuming that the dominant lobe at any given instant was whichever lobe Rosetta was directly above at that time (i.e. that most of the water in the part of the coma being measured by the DFMS at a given time was sublimating from that lobe), the D/H ratios of the lobes could be compared based on measurements of their HDO / $H_2O$. To circumvent any possible complications arising from changes in the detector's relative sensitivities over time, each D/H measurement of the small lobe was compared with the D/H measurement of the large lobe that was closest in time to it. The measurements in each pair were thus always made within 24 hours of each other. Border regions were avoided by excluding measurements taken within 10° longitude of each other.

However, the dominant lobe may not always be the lobe that Rosetta is directly above, since many other factors such as its distance from the nucleus and which side of 67P is illuminated by the Sun may also affect the amount of water contributed by each lobe to the mixture of gases eventually sampled by Rosetta at a certain position relative to the comet. An outgassing model recently developed by Combi et al. (2019) thus provides a superior alternative to the purely longitude-based definition of which lobe is dominant. This new model allows us to estimate the amount of water contributed by each individual lobe to any particular measurement made by Rosetta (Fig. 2 shows how the outgassing model defines the small and large lobes). We can thus refine our definition by instead considering a lobe to be dominant when its contribution to the DFMS signal is more than 80%.

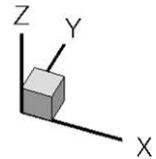
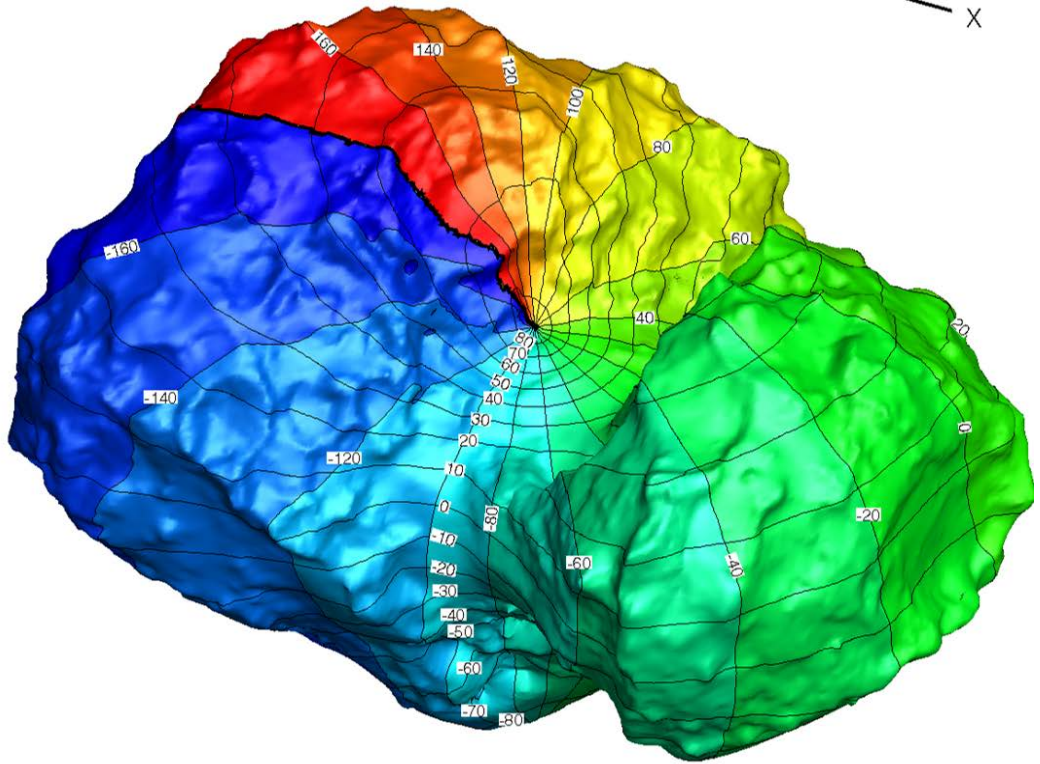

**Figure 1.** Shape model of comet 67P / Churyumov-Gerasimenko.

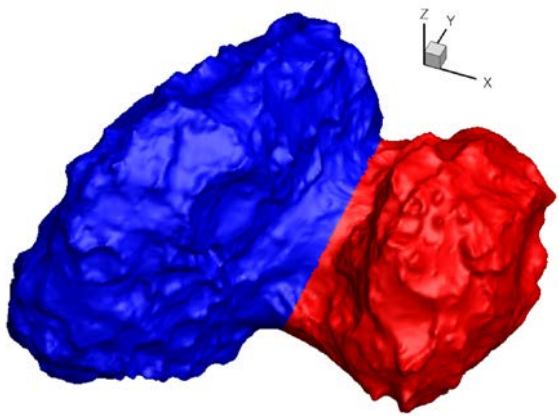

**Figure 2.** Definition of small (red) and large (blue) lobes as used by outgassing model.

A further enhancement can be made by accounting for the contributions from the non-dominant lobe as well. If the lobes differ in their D/H ratios, the overall D/H measured by the DFMS will then depend on the level of contribution from each lobe at the time of measurement and will change with Rosetta's position relative to the nucleus:

$$\frac{D}{H}_{overall}(t) = a_{small}(t) \cdot \frac{D}{H}_{small} + a_{big}(t) \cdot \frac{D}{H}_{big} \quad (1)$$

where $\frac{D}{H}_{overall}$ is the overall D/H ratio of 67P measured by the DFMS, $\frac{D}{H}_{small}$ and $\frac{D}{H}_{big}$ are the individual D/H ratios of the small and large lobes respectively and $a_{small}$ and $a_{big}$ are the respective contributions from the small and large lobes ($a_{small} + a_{big} = 1$). Taking two times, $t_1$ and $t_2$, which are close together to avoid any problems caused by changes in the instrument (e.g. in relative sensitivity) over time, Equation 1 can be rewritten in the form of two simultaneous equations and solved for $D/H_{small}$ and $D/H_{big}$:

$$\frac{D}{H}_{small} = \frac{\frac{D}{H}_{overall}(t_2) - \left(\frac{a_{big}(t_2)}{a_{big}(t_1)}\right)\frac{D}{H}_{overall}(t_1)}{a_{small}(t_2) - \left(\frac{a_{big}(t_2)}{a_{big}(t_1)}\right)a_{small}(t_1)} \quad (2)$$

$$\frac{D}{H}_{big} = \frac{\frac{D}{H}_{overall}(t_2) - \left(\frac{a_{small}(t_2)}{a_{small}(t_1)}\right)\frac{D}{H}_{overall}(t_1)}{a_{big}(t_2) - \left(\frac{a_{small}(t_2)}{a_{small}(t_1)}\right)a_{big}(t_1)} \quad (3)$$

## 3. Results & Discussion

Using the purely longitude-based approach, the mean $\frac{D/H_{small}}{D/H_{big}}$ ratio of the water in comet 67P was found to be 0.97 ± 0.08 (1σ error) based on 720 pairs of measurements of the small lobe and the large lobe made between October 2014 and August 2016 with both Rows A and B of the MCP / LEDA detector. These 720 individual values are shown plotted against their dates of measurement in Fig. 3, from which we can also observe that, despite the spread in the data, there were no discernible changes or trends over time. The data from Fig. 3 are further depicted in Fig. 4 in the form of a histogram. In our analysis, we have neglected the partial overlap of the $H_2^{17}O$ peak with the HDO peak in the mass spectra for m/z = 19 u/e, as the $H_2^{17}O$ signal was usually much smaller and could be ignored (Schroeder et al. 2019; Altwegg et al. 2014).

If we restrict our analysis solely to the 40 pairs of measurements from May and August 2016, which were the periods when Rosetta was at its closest to comet 67P (less than 20 km from the cometary nucleus), so as to minimise the influence of each lobe on measurements of the other as well as the effects of any extended sources of $H_2O$ which may be present (e.g. icy grains sublimating in the cometary coma), the difference between the small and large lobes becomes even smaller, with a mean $\frac{D/H_{small}}{D/H_{big}}$ ratio of 0.99 ± 0.08.

The same result was also obtained with the outgassing model-based approach of defining a lobe to be dominant only if its contribution to the gas density at the location of Rosetta exceeded 80%. Using this approach reduced our sample size to 271 pairs of measurements, as shown in Fig. 5, since the dominance of a lobe did not always correspond to longitude, as previously suspected, and there were many instances when the contribution from the large lobe was significant even though Rosetta was in orbit above the small lobe. However, this alternative approach still yields an average $\frac{D/H_{small}}{D/H_{big}}$ ratio of 0.99 ± 0.08.

Finally, Fig. 6 shows the outcome of our subsequent application of Equations 2 and 3 to the aforementioned 271 pairs of measurements, to further account for the influence of each lobe on measurements of the other. The average $\frac{D/H_{small}}{D/H_{big}}$ ratio of 67P, even after correcting each measurement for the contribution of the non-dominant lobe, is 0.99 ± 0.09.

The slightly larger error in the result derived with Equations 2 and 3 arises from the additional uncertainty introduced via the outgassing model's estimates of $a_{small}$ and $a_{big}$. Although the uncertainty in each of the model's individual predictions can reach as high as 30% (Combi et al. 2019), such random error, fortunately, is also largely reduced by our result being the mean of many measurements.

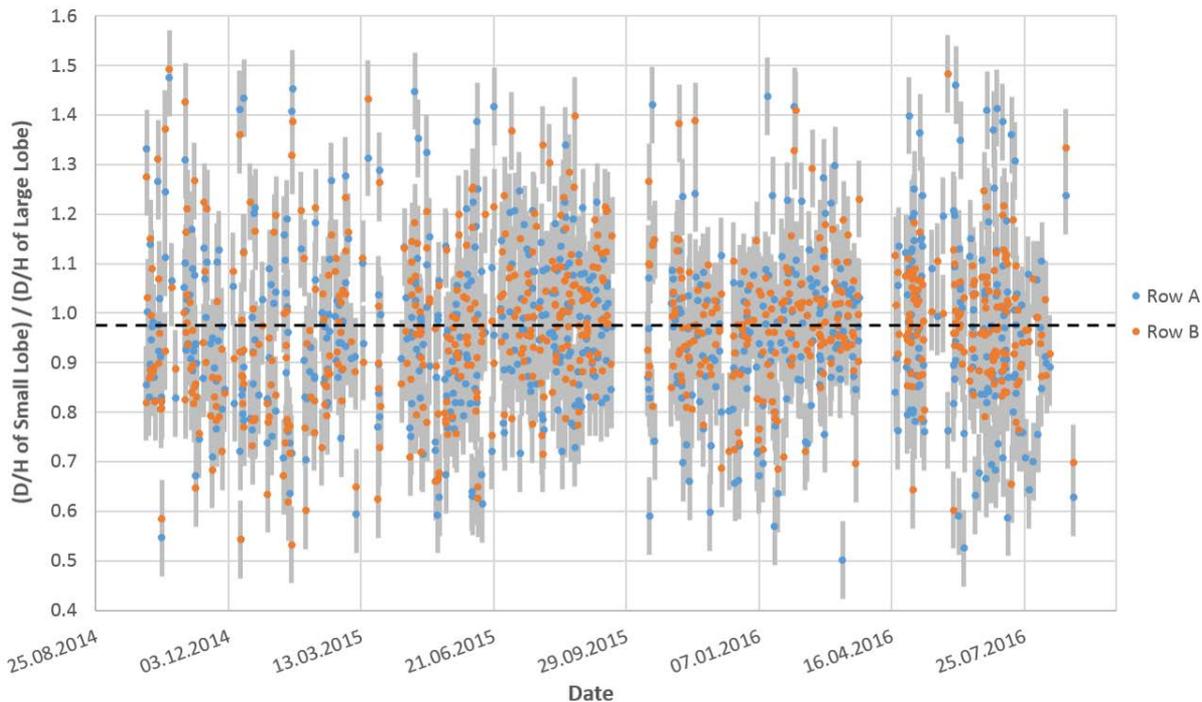

**Figure 3.** Ratio of D/H of small lobe over D/H of large lobe plotted against measurement date (longitude-based approach). The dotted black line represents the mean.

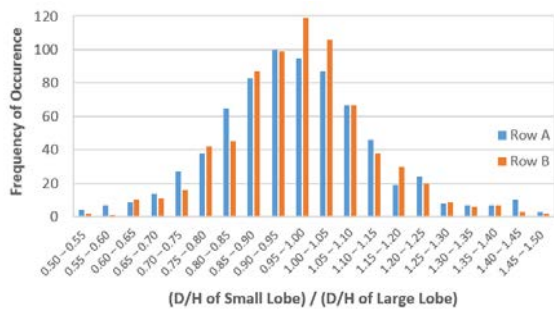

**Figure 4.** Histogram of individual measurements.

Many other factors (e.g. spacecraft orientation, duration of measurement, etc.) may also have affected the data, because the DFMS measured each mass separately. The dominant sources of error, however, remain the uncertainties in the detector gain (6%) and pixel gain (5%) (Schroeder et al. 2019). The statistical uncertainty, on the other hand, is an order of magnitude smaller, again as a consequence of our averaging over a large number of measurements.

The results from the various aforementioned methods with which we compared comet 67P's two lobes are all consistent with each other, within uncertainties. Furthermore, they reveal that any difference between the D/H ratios of water from the small and large lobes is much smaller than the uncertainty in the cometary D/H ratio of $(5.3 \pm 0.7) \times 10^{-4}$ itself (Altwegg et al. 2014, 2017). It thus appears that the D/H ratios of the two lobes are consistent with each other within $1\sigma$ error. This in turn supports the idea that, if the two lobes were formed separately, they were both formed in the same region, or at least in regions with similar D/H ratios, prior to their merger.

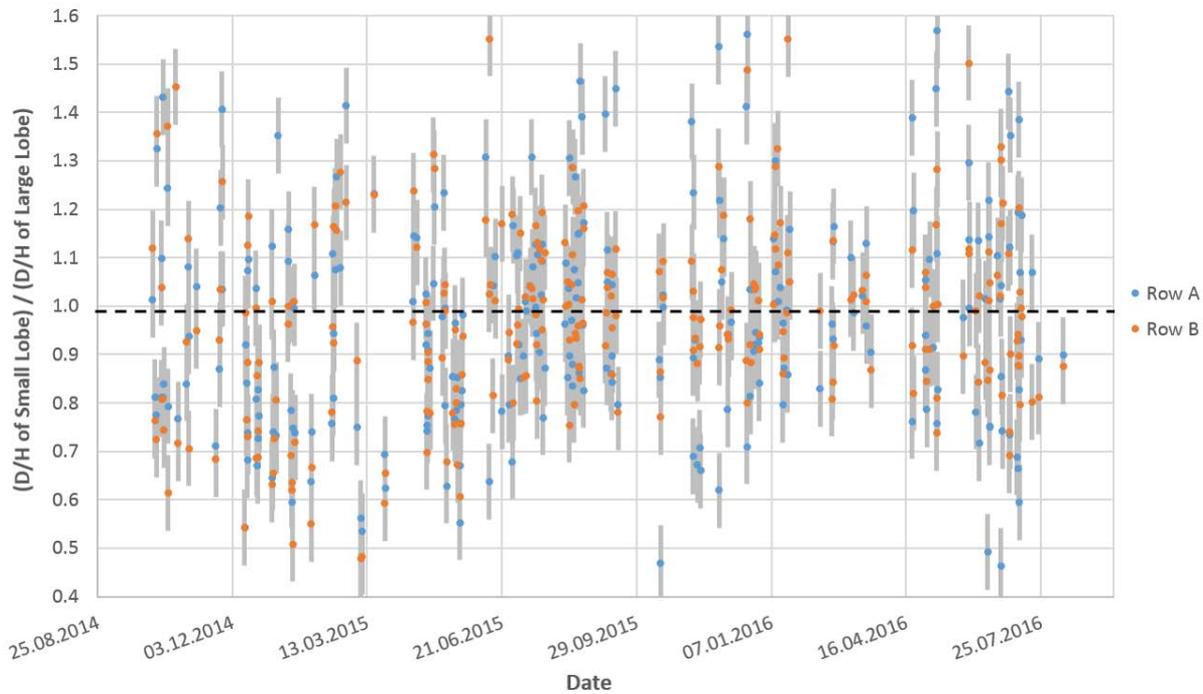

Figure 5. Ratio of D/H of small lobe over D/H of large lobe plotted against measurement date (outgassing model-based approach). The dotted black line represents the mean.

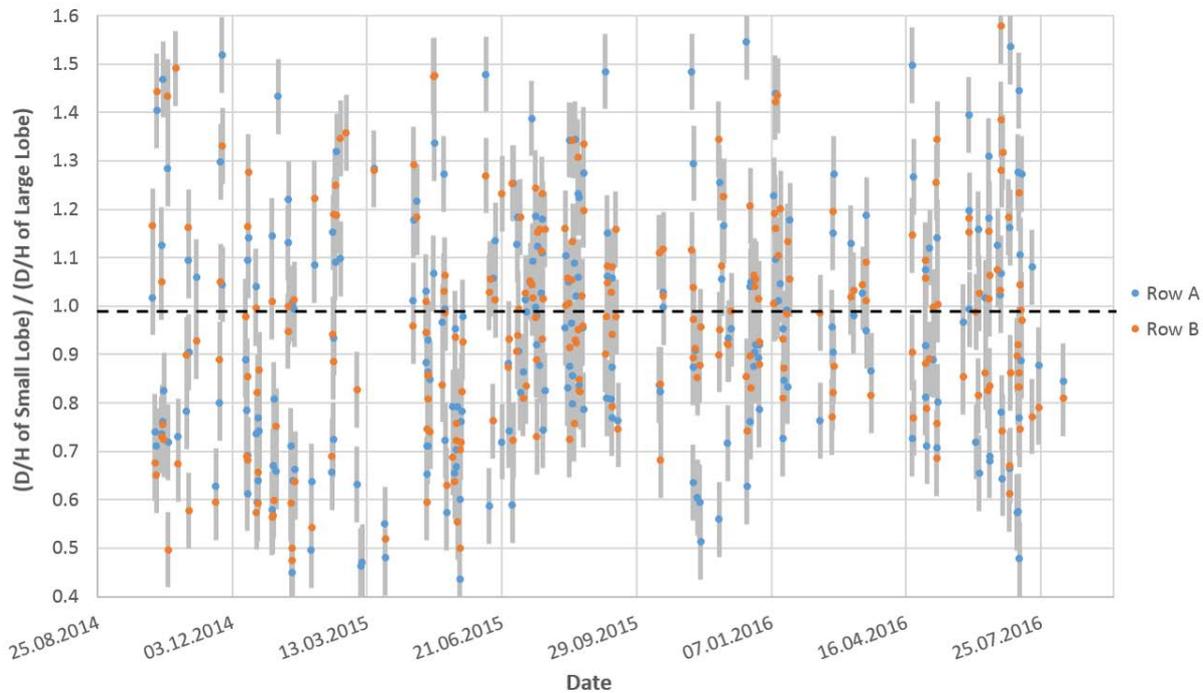

Figure 6. Ratio of D/H of small lobe over D/H of large lobe plotted against measurement date (outgassing model-based approach with additional corrections for contributions from non-dominant lobe).

## 4. Conclusions

We compared the Deuterium-to-Hydrogen ratios of the two lobes of comet 67P / Churyumov-Gerasimenko using a new outgassing model (Combi et al. 2019) to interpret the measurements of HDO and $H_2O$ obtained with data from the ROSINA Double Focusing Mass Spectrometer on board the Rosetta spacecraft, such that it was finally possible to make lobe-specific distinctions between them. No statistically significant evidence of any inhomogeneity in the D/H ratio was found, as any difference between the smaller and larger lobes of the comet, if it exists, is on the order of $\sim 1\%$, which is much smaller than the uncertainty in the cometary D/H ratio of $(5.3 \pm 0.7) \times 10^{-4}$ itself (Altwegg et al. 2014, 2017). This suggests that the lobes are homogeneous in their D/H ratios, which would be consistent with both lobes having been formed in the same region of the protoplanetary disc before they came together. A formation at similar radial distances in the protosolar nebula is compatible with the picture of a soft collision between both objects (Massironi et al. 2015).

## Acknowledgements

ROSINA would not have produced such outstanding results without the work of the many engineers, technicians and scientists involved in the mission, in the Rosetta spacecraft, and in the ROSINA instrument team over the last 20 years, whose contributions are gratefully acknowledged. Rosetta is a European Space Agency (ESA) mission with contributions from its member states and NASA. We acknowledge herewith the work of the entire ESA Rosetta team.

Work at the University of Bern was funded by the Canton of Bern, the Swiss National Science Foundation and the ESA PRODEX (PROgramme de Développement d'Expériences scientifiques) programme. Work at the Southwest Research Institute was supported by subcontract #1496541 from the Jet Propulsion Laboratory (JPL). Work at the Royal Belgian Institute for Space Aeronomy (BIRA-IASB) was supported by the Belgian Science Policy Office via PRODEX / ROSINA PRODEX Experiment Arrangement 90020. Work at the University of Michigan was funded by NASA under contract JPL-1266313.

## References

Albertsson T., Semenov D., Henning T., 2014, Chemodynamical deuterium fractionation in the early solar nebula: The origin of water on Earth and in asteroids and comets. Astrophys. J. 784, 39.

Altwegg K. et al., 2014, 67P / Churyumov-Gerasimenko, a Jupiter family comet with a high D/H ratio. Science 347, 1261952.

Altwegg K. et al., 2017, $D_2O$ and HDS in the coma of 67P / Churyumov-Gerasimenko. Phil. Trans. R. Soc. A 375: 20160253.

Balsiger H. et al., 2007, ROSINA - Rosetta Orbiter Spectrometer for Ion and Neutral Analysis. Space Science Reviews, 128, 745-801.

Bieler A. et al., 2015, Comparison of 3D kinetic and hydrodynamic models to ROSINA-COPS measurements of the neutral coma of 67P / Churyumov-Gerasimenko. A&A 583, A7.

Biver N. et al., 2016, Isotopic ratios of H, C, N, O, and S in comets C/2012 F6 (Lemmon) and C/2014 Q2 (Lovejoy). A&A 589, A78.

Brasser R., Morbidelli A., 2013, Oort cloud and scattered disc formation during a late dynamical instability in the solar system. Icarus 225, 40-49.

Ceccarelli C. et al., 2014, Deuterium fractionation: the Ariadne's thread from the pre-collapse phase to meteorites and comets today. Protostars and Planets VI.

Combi M. et al., 2019, The Surface Distributions of the Production of the Major Volatile Species, $H_2O$, $CO_2$, CO and $O_2$,

from the Nucleus of Comet 67P / Churyumov-Gerasimenko throughout the Rosetta Mission as Measured by the ROSINA Double Focusing Mass Spectrometer. Icarus (in preparation).

Duncan M. J., Levison H. F., 1997, A disk of scattered icy objects and the origin of Jupiter-family comets. Science 276, 1670-1672.

Furuya K., Aikawa Y., Nomura H., Hersant F., Wakelam V., 2013, Water in protoplanetary disks: Deuteration and turbulent mixing. Astrophys. J. 779, 11.

Geiss J., Gloeckler G., 1998, Abundances of Deuterium and Helium-3 in the protosolar cloud. Space Sci. Rev. 84, 239-250.

Geiss J., Reeves H., 1981, Deuterium in the solar system. A&A 93, 189-199.

Hartogh P. et al., 2011, Ocean-like water in the Jupiter-family comet 103P / Hartley 2. Nature 478, 218-220.

Hässig M. et al., 2013, ROSINA / DFMS capabilities to measure isotopic ratios in water at comet 67P / Churyumov-Gerasimenko. Planet. Space Sci. 84, 148-152.

Kavelaars J. J., Mousis O., Petit J.-M., Weaver J. A., 2011, On the formation location of Uranus and Neptune as constrained by dynamical and chemical models of comets. Astrophys. J. 734, L30.

Keller H. U. et al., 2007, OSIRIS – The Scientific Camera System Onboard Rosetta. Space Sci. Rev. 128, 433-506.

Levison H. F., Duncan M. J., Brasser R., Kaufmann D. E., 2010, Capture of the Sun's Oort cloud from stars in its birth cluster. Science 329, 187-190.

Linsky J. L. et al., 2006, What is the total deuterium abundance in the local galactic disk? Astrophys. J. 647, 1106-1124.

Lis D. C. et al., 2019, Terrestrial deuterium-to-hydrogen ratio in water in hyperactive comets. A&A 625, L5.

Mahaffy P. R. et al., 1998, Galileo Probe measurements of D/H and $^3$He / $^4$He in Jupiter's atmosphere. Space Sci. Rev. 84, 251-263.

Massironi M. et al., 2015, Two independent and primitive envelopes of the bi-lobate nucleus of comet 67P. Nature 526, 402-405.

Nesvorný D. et al., 2018, Bi-lobed Shape of Comet 67P from a Collapsed Binary. AJ 155, 246.

Rickman H. et al., 2015, Comet 67P / Churyumov-Gerasimenko: Constraints on its Origin from OSIRIS Observations. A&A 583, A44.

Schroeder I. et al., 2019, The $^{16}$O/$^{18}$O ratio in water in the coma of comet 67P / Churyumov-Gerasimenko measured with the Rosetta / ROSINA double-focusing mass spectrometer. A&A, Rosetta special issue 2 (Accepted, DOI: https://doi.org/10.1051/0004-6361/201833806).

Visser R., Doty S. D., van Dishoeck E. F., 2011, The chemical history of molecules in circumstellar disks. II. Gas-phase species. A&A 534, A132.

## Appendix A: Outgassing Model

Combi et al. (2019) used an analytic model which incorporates DFMS measurements, solar illumination conditions, spacecraft position and the geometry of the nucleus surface to derive potential surface-activity maps for several volatile species. The potential activity of the k$^{th}$ triangle on 67P's surface is $f_k = \sum_{j=1}^{N} x_j Y_j(\vartheta_k, \varphi_k)$, where $Y_j$ is the j$^{th}$ spherical harmonics function, $x_j$ is its coefficient and $\vartheta_k$ and $\varphi_k$ are the colatitude and longitude of the centre of the k$^{th}$ triangle respectively. An *M x N* matrix, *C*, establishes the connection between the location of the spacecraft, the extent to which the nucleus is illuminated by the sun and the geometric projection of each surface facet of the nucleus, as seen by Rosetta:

$$C_{ij} = \frac{1}{R_{AU}^{\beta}} \left( \sum_{k=1}^{N_{faces}} \frac{(g(\Theta_k))_i S_k (\cos(\alpha_k))_i}{r_k^2} Y_j(\vartheta_k, \varphi_k) \right) \quad (A1)$$

where $N_{faces}$ is the total number of triangles in the nucleus surface mesh, $\Theta_k$ is their solar zenith angle, $g$ is defined as in Bieler et al. (2015) by $g(\Theta_k) = max(a_{night}, \cos \Theta_k)$ and $a_{night}$ defines a lower limit for the surface activity on the night side relative to the day side. In Combi et al. (2019), $a_{night}$ is set to 0.02 (i.e. the minimum night-side activity is assumed to be 2% of the day-side maximum). The surface area of the $k^{th}$ triangle is $S_k$, while $r_k$ and $\alpha_k$ are the distance and angle between the spacecraft and the outward normal of the $k^{th}$ triangle respectively. $\beta$ describes the long-term variations in the production rate with heliocentric distance. The coefficients of the spherical harmonics can be obtained by solving the least squares problem $\min_{x \in R^N} \|Cx - d\|_2^2$ with the constraint that the surface activity $f_k$ is positive definite, where $d$ is a vector with a length of $M$ elements that contains the DFMS data and $x$ is a vector of length $N$ containing the undetermined coefficients of the spherical harmonics. As Combi et al. (2019) computed the spherical harmonic coefficients $x$, the respective gas density contributions from the large lobe $d_b$ and small lobe $d_s$ can be obtained via the matrix multiplication $d_b = \boldsymbol{B}x$ and $d_s = \boldsymbol{S}x$, where:

$$B_{ij} = \frac{1}{R_{AU}^{\beta}} \left( \sum_{k=1}^{N_{big}} \frac{(g(\Theta_k))_i S_k (\cos(\alpha_k))_i}{r_k^2} Y_j(\vartheta_k, \varphi_k) \right) \quad (A2)$$

$$S_{ij} = \frac{1}{R_{AU}^{\beta}} \left( \sum_{k=1}^{N_{small}} \frac{(g(\Theta_k))_i S_k (\cos(\alpha_k))_i}{r_k^2} Y_j(\vartheta_k, \varphi_k) \right) \quad (A3)$$

The $\boldsymbol{B}$ matrix sums over triangles on the surface of the large lobe, while $\boldsymbol{S}$ sums over triangles on the small lobe.